\documentclass[11pt,twoside]{article}

\usepackage{asp2014}

\pdfpageattr {/Group << /S /Transparency /I true /CS /DeviceRGB>>}

\aspSuppressVolSlug
\resetcounters

\begin{document}

\title{All of the Sky: HEALPix Density Maps of Gaia-scale Datasets
       from the Database to the Desktop}

\author{M.~B.~Taylor,$^1$ G.~Mantelet,$^2$ and M.~Demleitner,$^2$
\affil{$^1$H.~H.~Wills Physics Laboratory, University of Bristol, U.K.}
\affil{$^2$Astronomisches Rechen-Institut, ZAH, Universit\"{a}t Heidelberg, Germany}}

\paperauthor{M.~B.~Taylor}{m.b.taylor@bristol.ac.uk}{0000-0002-4209-1479}{University of Bristol}{School of Physics}{Bristol}{Bristol}{BS8 1TL}{U.K.}
\paperauthor{G.~Mantelet}{gmantele@ari.uni-heidelberg.de}{}{University of Heidelberg}{Astronomisches Rechen-Institut}{Heidelberg}{Baden-W\"urttemberg}{69120}{Germany}
\paperauthor{M.~Demleitner}{msdemlei@ari.uni-heidelberg.de}{}{University of Heidelberg}{Astronomisches Rechen-Institut}{Heidelberg}{Baden-W\"urttemberg}{69120}{Germany}

\begin{abstract}
The Gaia Archive provides access to observations of around
a billion sky sources.
The primary access to this archive is via TAP services such as GACS
and ARI-Gaia,
which allow execution of SQL-like queries against a large remote database
returning a result set of manageable size for client-side use.
Such services are generally used for extracting relatively small source
lists according to potentially complex selection criteria.
But they can also be used to obtain statistical information
about all, or a large fraction of, the observed sources
by building histogram-like results.

We examine here the practicalities of producing and consuming
all-sky HEALPix weighted density maps in this way for Gaia and
other large datasets.
We present some modest requirements on TAP/RDBMS services
to enable such queries, and discuss visualisation and serialization
options for the results including some new capabilities in recent
versions of TOPCAT.
\end{abstract}

\section{Introduction}

The primary data access for {\em Gaia} \citep{2016arXiv160904172G}
and several other past and upcoming
large-scale surveys is via Table Access Protocol (TAP)
services that allow users to
execute SQL-like queries against a large remote database.
This model of bringing the computation to the data is enforced by the
size of these datasets; client-side transportation,
storage and processing of the whole dataset
is for most purposes impractical or at least highly inefficient.

The remote database engines are typically powerful and can perform
fast execution of complex queries.  Where the desired result is some
kind of source list of limited size, filtered by criteria such
as sky position or photometry down to no more than a few thousand
or maybe million objects, selection on source criteria works well.
But where the requirement is to sample all or a large fraction
of the sources in a catalogue in order to obtain statistical
information about all or large regions of the sky,
the model of retrieving source lists breaks down,
since results with very large row counts are disallowed by the service
or simply unwieldy to transport to and process at the client.

It is however possible to calculate in the database histograms
representing statistical aggregations of all or many data rows.
By binning into a tessellating grid of sky tiles, queries can
produce weighted or unweighted sky density maps representing
source density or other statistical quantities by sky position.
Such queries can be executed in reasonable amounts of time
and provide result sets small enough to be transported to the
client for examination and analysis.

\section{Tiling Scheme}

Various sky tiling schemes exist, including HTM, Q3C, and HEALPix.
We favour the NESTED variant of HEALPix \citep{2005ApJ...622..759G}
which has a number of advantages for this application, including
the facts that tiles have equal area, facilitating density map analysis,
and that simple SQL-friendly arithmetic (integer division)
can be used to degrade pixel index to a lower resolution.
The HEALPix grid at order $N$ defines tiles with indices
in the range $[ 0, 12 \times 4^{N} )$.
A sky position within tile $i$ at order $N$
falls within tile $i/4^{N-M}$ at a lower order
(coarser resolution) $M$.

\section{Service Requirements}

The following items must be in place for end-users to be able to
construct and use customised weighted or unweighted all-sky
density maps for catalogues that would be impractical to download:
\begin{description}
\item[SQL-like access to source catalogue:]
Public datasets are increasingly exposed via the Virtual Observatory
protocol TAP (Table Access Protocol), allowing remote execution of
ADQL (SQL-like) queries.
\item[HEALPix column or function:]
{\em Either\/} the table must have a column giving the index
of the HEALPix tile in which the source position falls,
{\em or\/} a User-Defined Function must exist that
can calculate tile index for each row (e.g.\ from RA, Dec columns).
Most existing TAP services do not currently provide this,
but the ARI-Gaia and DaCHS TAP services have introduced such a UDF
{\em (this work)\/}:
\begin{quote}
   {\tt ivo\_healpix\_index(order, ra, dec)}
\end{quote}
An order-12 HEALPix index is also buried in bits 36--63 of the
Gaia {\tt source\_id} column and can be extracted by integer division.
\item[GROUP BY query:]
An SQL query of the form
\begin{quote}
   {\tt SELECT {\sl (agg-func)}
        FROM {\sl (table)}
        GROUP BY {\sl (healpix-index)}}
\end{quote}
calculates the sky map, returning one row per populated sky pixel.
The aggregate function defines the weighting
(e.g.\ {\tt COUNT(*)} gives unweighted source density,
{\tt AVG(x)} gives the mean value of column or expression {\tt x})
and a {\tt WHERE} clause can optionally be added
to restrict the selection of sources.
\item[Query limits:]
Limits on query execution time and output size must accommodate
execution of these aggregating queries.
They typically take very roughly an hour per billion rows,
which is long but not unfeasibly so.
Million-row outputs are a convenient size for visualisation
(HEALPix order 8 has 786\,432 tiles)
though finer or coarser resolutions can also be useful.
Some TAP services impose limits on execution time or output row count
that can preclude these queries.
\item[Semantic markup of HEALPix output:]
An undocumented convention exists for serialization of HEALPix maps
in FITS files, but not for VOTable, which is the standard output
format for TAP.  Discussion is ongoing in the IVOA about how best
to do this.
\end{description}


\begin{figure}
\plotone{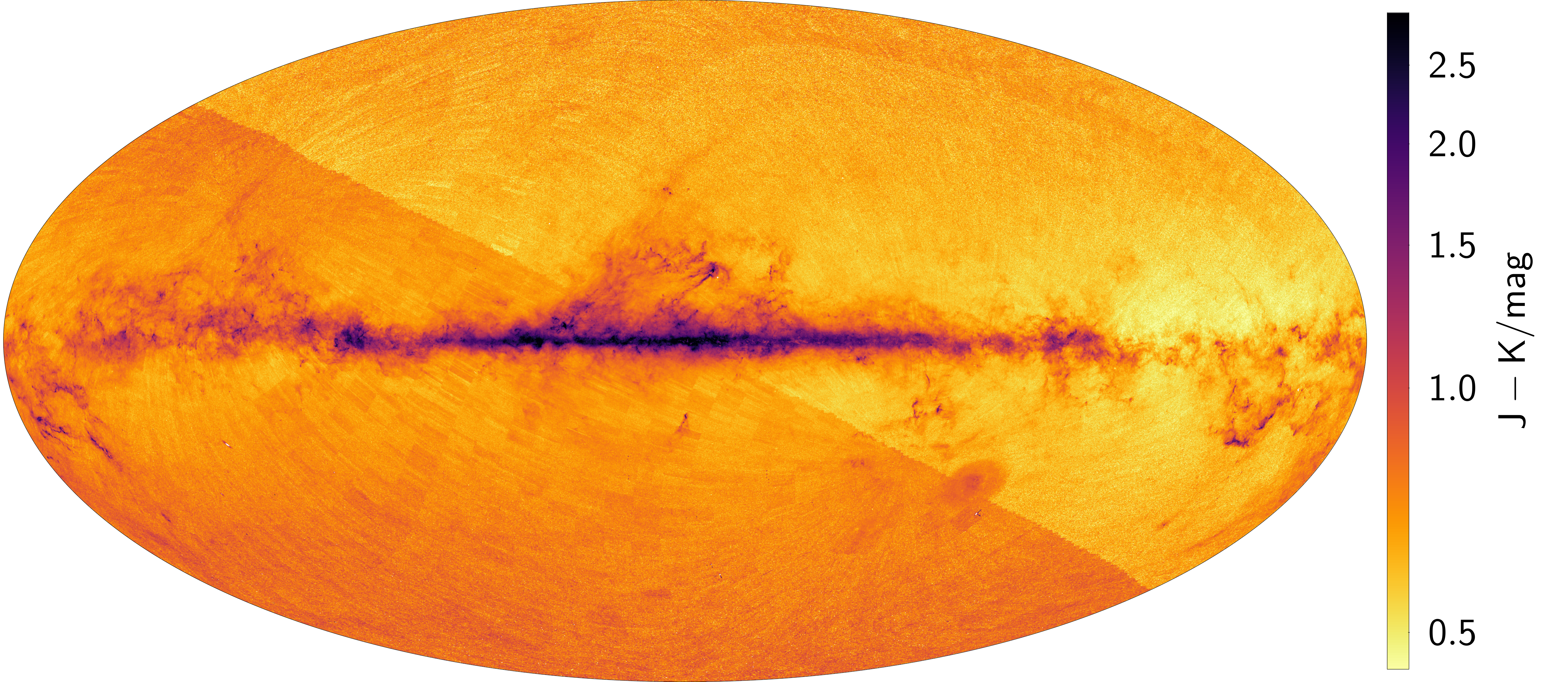}
\caption[figure 1]{ \label{P1-31:2mass}
$J-K$ colour for 2MASS point sources, using the query:
``{\tt\footnotesize
  SELECT ivo\_healpix\_index(9,raj2000,dej2000) AS hpx9,
         AVG(jmag-kmag) AS j\_k 
  FROM twomass.data
  WHERE qflg LIKE 'A\_A' AND cflg LIKE '0\_0' AND xflg = '0'
  GROUP BY hpx9}''.
The proposed User-Defined Function is used to calculate
HEALPix index from sky position.
The upper right half of the image used
the {\tt WHERE} clause above,
which selects only
sources with good J/K photometry, while the lower left includes all sources
(no {\tt WHERE} clause).
With the custom selection
the image is cleaner and the values are lower
on average, though not uniformly over the sky.
This query took 16/39 minutes to scan 163/471 million rows using the
GAVO DC TAP service.  Plot by STILTS.
}
\end{figure}

\begin{figure}
\plotone{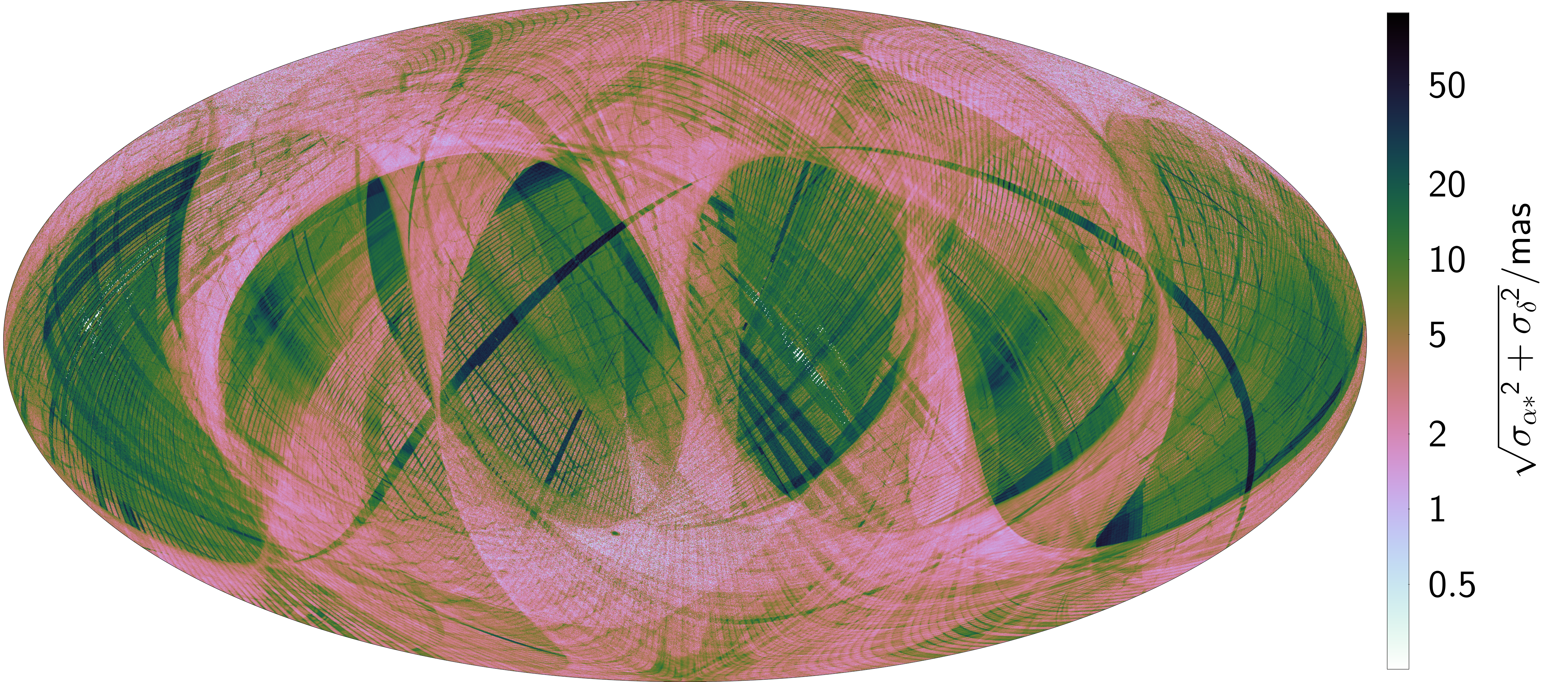}
\caption[figure 2]{ \label{P1-31:poserr}
Mean isotropic positional error of Gaia DR1 sky positions,
using the query:
``{\tt\footnotesize
  SELECT source\_id/2199023255552 AS hpx9,
         AVG(SQRT(ra\_error*ra\_error+dec\_error*dec\_error)) AS pos\_error
  FROM gaia.dr1
  GROUP BY hpx9}''.
The HEALPix index is recovered from the Gaia {\tt source\_id}
column using integer division.
This query took 70 minutes to scan 1.1 billion rows using the
GAVO DC TAP service.  Plot by STILTS.
}
\end{figure}

\section{Analysis in TOPCAT and STILTS}

Recent releases of the TOPCAT/STILTS table analysis
suite \citep{2005ASPC..347...29T} include new
features for working with HEALPix maps.
Tables with an implicit or explicit HEALPix index column
can be visualised interactively or exported to bitmapped or
vector graphics files.
They can be displayed within TOPCAT's Sky Plot window which offers
interactive adjustment of colour maps and grid resolution,
pan/zoom navigation, a choice of sky projections and coordinate systems,
and the option to overlay multiple plots of different types.
Figures \ref{P1-31:2mass} and \ref{P1-31:poserr} show examples
plotted by STILTS.
There are also new capabilities to generate HEALPix maps on
the client side from local source catalogues and a number of
HEALPix-related functions added to the expression language.
Since HEALPix maps are tables, these tools can be used to
analyse and manipulate them in general, non-visual ways too,
for instance calculating statistics and performing joins.


\section{Conclusions}

An all-sky or wide-field view of quantities aggregated from
a large catalogue can sometimes reveal large scale features
or trends in astronomical or instrumental behaviour that would
be difficult to discern from other data products.
Source density maps are the most obvious application,
but there are numerous other possibilities.

Although some data centers (including the ESA and ARI Gaia archives)
offer for download various pre-calculated all-sky maps in graphical
or tabular form, it is often useful for end-users to construct their
own, for instance applying custom source selections or
weighting functions not foreseen by data centers.
Two examples are given in the figures.

We show that this is feasible using TAP services given certain
modest requirements.
Although this technique is not novel, the lack of required features
in most existing TAP services indicate that it is not widely practised.

To enable more widespread use of this technique, we recommend that
TAP services should make available the User-Defined Function
{\tt ivo\_healpix\_index(order, ra, dec)},
and should also consider the case of sky map creation
when setting query timeout and row output limits.
We also encourage the IVOA to standardise the representation of
HEALPix tile indices in VOTables.

\acknowledgements This project has received funding from the EU FP7-SPACE-2013-1 grant 606740 (GENIUS), the UK's STFC grant ST/M000907/1 (Gaia CU9), and the BMBF grant 05A11VH3 (GAVO).  It has made use of data from the ESA mission {\em Gaia} processed by DPAC, and from the UMass/IPAC/CalTech project 2MASS.

\bibliography{P1-31}

\end{document}